\documentclass[10pt,twocolumn,letterpaper]{article}

\usepackage{geometry}
\usepackage[prologue,table]{xcolor}    
\usepackage{balance}
\usepackage{mathtools}
\usepackage{graphicx}
\usepackage[labelfont=bf]{caption}
\usepackage{subcaption}
\usepackage{color}
\usepackage{datetime}
\usepackage{xurl}
\usepackage{hyperref}
\usepackage{tabularx}
\usepackage{chngcntr}
\usepackage{apptools}
\usepackage{amssymb,amsmath,amsthm}
\usepackage{pifont}
\usepackage{xspace}
\usepackage{comment}
\usepackage{multirow}
\usepackage{wrapfig}
\usepackage{verbatim}
\usepackage{authblk}

\usepackage[show]{chato-notes}
\newcommand{\spara}[1]{\vspace{2mm}\noindent\textbf{#1} }

\definecolor{cpur}{rgb}{0.50,0.00,0.25}
\definecolor{red}{rgb}{1.00,0.00,0.00}
\definecolor{darkgreen}{rgb}{0.078,0.667,0.016}
\definecolor{gray}{rgb}{0.5,0.5,0.5}

\newcommand{\cb}{\textcolor{blue}}

\newcommand{\cred}{\textcolor{red}}

\newcommand{\para}[1]{\noindent \textbf{#1}}

\newcommand{\SYS}{LiveGraph\xspace}
\newcommand{\LOG}{TEL\xspace}
\newcommand{\LOGs}{TELs\xspace}

\newcolumntype{Y}{>{\centering\arraybackslash}X}

\makeatletter
\newcommand\footnoteref[1]{\protected@xdef\@thefnmark{\ref{#1}}\@footnotemark}
\makeatother
\graphicspath{{figures/}}

\allowdisplaybreaks[1] 
\hypersetup{draft}
\AtAppendix{\counterwithin{theorem}{section}} 

\begin{document}


\title{\vspace{-50pt}\SYS: A Transactional Graph Storage System with Purely Sequential Adjacency List Scans}

\author[1]{Xiaowei Zhu}
\author[1]{Guanyu Feng}
\author[2]{Marco Serafini}
\author[3]{Xiaosong Ma}
\author[1]{Jiping Yu}
\author[1]{Lei Xie}
\author[3]{Ashraf Aboulnaga}
\author[1,4]{Wenguang Chen\thanks{Wenguang Chen is the corresponding author.}
}
{
\makeatletter
\renewcommand\AB@affilsepx{, \protect\Affilfont}
\makeatother
\affil[1]{\normalsize Tsinghua University}
\affil[2]{University of Massachusetts Amherst}
\affil[3]{Qatar Computing Research Institute}
\affil[4]{Beijing National Research Center for Information Science and Technology\\}
}

\affil[1]{\texttt{\{zhuxiaowei,cwg\}@tsinghua.edu.cn; \{fgy18,yjp19,xie-l18\}@mails.tsinghua.edu.cn;}}
\affil[2]{\texttt{marco@cs.umass.edu;}}
\affil[3]{\texttt{\{xma,aaboulnaga\}@hbku.edu.qa}}

\date{}

\maketitle

\begin{abstract}
The specific characteristics of graph workloads make it hard to design a one-size-fits-all graph storage system.
Systems that support transactional updates use data structures with poor data locality, which limits the efficiency of analytical workloads or even simple edge scans.
Other systems run graph analytics workloads efficiently, but cannot properly support transactions.

This paper presents \SYS, a graph storage system that outperforms both the best graph transactional systems \emph{and} the best solutions for real-time graph analytics on fresh data.
\SYS achieves this by ensuring that adjacency list scans, a key operation in graph workloads, are \emph{purely sequential}: they never require random accesses even in presence of concurrent transactions.
Such pure-sequential operations are enabled by combining a novel graph-aware data structure, the \textit{Transactional Edge Log (\LOG)}, with a concurrency control mechanism that leverages \LOG's data layout.
Our evaluation shows that \SYS significantly outperforms state-of-the-art (graph) database solutions on both transactional and real-time analytical workloads.
\end{abstract}


\section{Introduction}
\label{sec:intro}

Graph data is one of the fastest-growing areas in data management:
applications performing graph processing and graph data management are predicted to double annually through 2022~\cite{gartner_report_top_10_data_analytics_tech}.
Applications using graph data are extremely diverse.
We can identify
two broad classes of graph workloads with different requirements: \emph{transactional graph data management} and \emph{graph analytics}.

Transactional graph data management workloads continuously update and query single vertices, edges, and adjacency lists.\footnote{We call a workload ``transactional'' if it consists of simple read/write operations that must be interactive and require very low latency, 
regardless of whether the transactions
access only one object or multiple objects atomically.}
Facebook, for example, stores posts, friendship relationships, comments, and other critical data in a graph format~\cite{armstrong2013linkbench,bronson2013tao}.
Write transactions incrementally update the graph, while read transactions are localized to edges, vertices, or the neighborhood of single vertices.
These applications require a graph storage system to have very low latency and high throughput,
be it a key-value store, a relational database management system, or a
specialized graph database system.
The system must also have classic transactional features: concurrency control to deal with concurrent updates and durability to persist updates.

Graph analytics tasks run on a consistent read-only graph snapshot and their performance highly depends on efficient scans of the neighborhood of a vertex (i.e., the adjacency list of the vertex).
A particular class of analytics, \emph{real-time} analytics on \emph{fresh dynamic} graph data, is becoming increasingly important.
For example,
consider recommendations, where a website needs to find relevant products/connections based on users' properties and most recent interactions, which reflect their interests at the moment. 
Other applications in this class include privacy-related data governance (where ``expired'' data needs to be excluded from analytics for GDPR compliance),
finance (where financial institutions establish if groups of people connected through common addresses, telephone numbers, or frequent contacts are issuing fraudulent transactions), or systems security (where monitoring systems detect whether an attacker has performed a sequence of correlated steps to penetrate a system). 

It is increasingly attractive to have a graph storage system that simultaneously supports both transactional and (real-time) analytical workloads.
Unfortunately, common data structures adopted separately in the two worlds do not fare well when crossing into unfamiliar
territory.

Data structures used in state-of-the-art DBMSs and key-value stores do not support well adjacency list scans, a crucial operation in graph analytics and graph database queries. 
More specifically, popular structures such as B+ trees and Log-Structured Merge Trees (LSMTs) yield significantly worse performance in graph analytics than graph-aware data structures like Compressed Sparse Rows (CSR)~\cite{saad2003iterative}.
We performed micro-benchmarks and a micro-architectural evaluation comparing alternative data structures for storing graph data, and in particular adjacency lists.
The results show that 
contiguous in-memory storage of adjacency lists not only improves caching efficiency, but also allows better speculation and prefetching, 
reducing both memory access costs and the number of instructions executed.

At the other end of the spectrum, analytical graph engines often use sequential memory layouts for adjacency lists like CSR.
They feature efficient scans but do not support high-throughput, low-latency concurrent transaction processing. 
In fact, most existing graph engines do not target mutable graphs at all. 
Adding concurrency control to deal with concurrent updates is not straightforward.
The concurrency control algorithm is on the critical path of every operation and thus directly impacts the performance of adjacency list scans.
It should not disrupt otherwise sequential scans with random accesses or a complex execution flow.
There has been much recent work on in-memory concurrency control and transactional support for relational data~\cite{faleiro2015rethinking, larson2011high, lim2017cicada, neumann2015fast, tu2013speedy, wu2017empirical}, but none of the existing studies has specifically targeted the unique requirements of graph workloads.

This paper is a first step towards filling this gap.
It proposes \SYS, a graph storage system supporting both transactional and (real-time) analytical workloads.
A key design goal of \SYS is to ensure that adjacency list scans are \emph{purely sequential}, that is, they never require random accesses even in the presence of concurrent transactions.
To this end, we \emph{co-design} the system's graph-aware data structure and its concurrency control algorithm.
\SYS stores adjacency lists in a new data structure called the \emph{Transactional Edge Log} (\LOG).
The \LOG combines multi-versioning with a sequential memory layout.
The concurrency control of \SYS leverages the cache-aligned timestamps and counters of the \LOG to preserve the sequential nature of scans even in the presence of concurrent transactions.
It is an efficient yet simple algorithm whose regular execution flow enables speculation and prefetching.

Our evaluation compares \SYS with several state-of-the-art systems, including specialized graph databases, graph database solutions built on key-value stores or traditional RDBMSes, and graph engines. 
Results demonstrate that \SYS outperforms the current leaders at their specialty, in particular outperforming Facebook's RocksDB~\cite{rocksdb} by up to 7.45$\times$ using Facebook's social graph benchmark~\cite{armstrong2013linkbench}.
In addition, \SYS dramatically outperforms (up to $36.4 \times$ better than the runner-up) all systems that have implemented the LDBC SNB interactive workload~\cite{erling2015ldbc}, ingesting updates and performing real-time analytics queries concurrently.  
We further perform micro-benchmarking and extensive profiling to understand the performance differences. 
Finally, \SYS allows 
lower end-to-end processing time by conducting in-situ iterative graph analytics (like PageRank) on its latest snapshot, as the expensive ETL cost can now be eliminated.

This paper is an extended version of~\cite{LiveGraph-pvldb}.
In Section~\ref{sec:microbenchmark-perf}, we motivate the importance of sequential adjacency list scans.
We present the design of \SYS in Sections~\ref{sec:representation}--\ref{sec:storage}.
Section~\ref{sec:eval} presents our evaluation results, Section~\ref{sec:relwork} presents related work, and Section~\ref{sec:conclusions} concludes.

\section{Purely Sequential Scans}
\label{sec:microbenchmark-perf}

A key design choice of \SYS is ensuring purely sequential adjacency list scans: scans should never entail random accesses.
Before introducing the details of \SYS, we motivate why purely sequential adjacency list scans are important.
We use single-threaded micro-benchmarks and micro-architectural analysis to compare different commonly used data structures and quantify the advantage of a sequential memory layout.
Then, we discuss how concurrency control algorithms introduce additional complexity in the form of random accesses and branching.

\subsection{The Benefits of Sequential Edge Storage}
Adjacency lists contain the key topological information in a graph.
Full or partial scans of these lists are fundamental operations in graph workloads, from simple queries to full-graph analytics. 
Graph storage must balance fast scans with efficient edge insertions, which are frequent in graph writes~\cite{bronson2013tao,armstrong2013linkbench}.
In the following, we compare the scan performance of different data structures used for graph storage.

\begin{table}
\center
\caption{Adjacency list scan properties of different data structures. $N$ is the size of the tree.\label{tab:comparison}}
\resizebox{\columnwidth}{!}{
\begin{tabular}{c|c c|c}
\hline
Cost & \multicolumn{2}{c|}{Seek} & Scan (per edge) \\
\hline
B+ Tree & $O(\log N)$ & random & sequential w.\ random\\
LSMT & $O(\log N)$ & random & sequential w.\ random\\
Linked List & $O(1)$ & random & random \\
CSR & $O(1)$ & random & sequential \\
\hline
\LOG & $O(1)$ & random & sequential \\
\hline
\end{tabular}
}
\end{table}

\spara{Graph data representations.}
Graph data consists of two types of objects: vertices and edges.

The CSR representation
consists of two arrays, 
the first storing the adjacency lists of all vertices as sequences of destination vertex IDs, while the second storing pointers to the first array, indexed by source vertex ID.
CSR is very compact, leading to a small storage footprint, reduced memory traffic, and high cache efficiency.
Also, unlike most other data structures, it enables pure sequential adjacency list scans.
These properties make it a top choice for graph engines~\cite{nguyen2013galois,gonzalez2012powergraph,zhu2016gemini}, which target read-only analytical workloads.
On the flip side, it is immutable, making it unsuitable for
dynamic graphs or 
transactional workloads.

Linked list 
is an intuitive choice for adjacency lists and is used by Neo4j~\cite{neo4j}, a popular transactional graph database.
It easily supports edge insertions but suffers from random accesses during scans when traversing through pointers.

Other state-of-the-art graph stores adopt general-purpose data structures such as 
the B+ tree 
and the LSMT (Log-Structured Merge-Tree). 
The adjacency list is represented as a single sorted collection of edges, whose unique key is a $\langle$\texttt{src},\texttt{dest}$\rangle$ vertex ID pair.

In this work, we propose a new data structure, the Transactional Edge Log (\LOG), which simultaneously allows sequential adjacency list scans and fast edge insertion. 
Unlike existing structures used in graph systems, it features purely sequential, yet mutable, edge storage. 
For this discussion, it suffices to say that edges in the same adjacency list are stored sequentially in contiguous blocks with empty slots at the tail.
Edge insertions and updates are appended at the tail of the block until it fills up, at which point the \LOG is upgraded to a larger block. 
Like CSR, \LOG has purely sequential adjacency list scans. 

Table~\ref{tab:comparison} compares major operation complexities of the aforementioned data structures.
Each adjacency list scan consists of a one-time \emph{seek} operation, which locates the first edge of the adjacency list, followed by 
an \emph{edge scan}, a sequence of edge accesses. 
Note that the initial seek cost often cannot be amortized, especially considering that most real-world graphs exhibit power-law degree distributions so most vertices have few edges.
For a more detailed comparison, we refer to a comprehensive survey of data structures used in graph databases~\cite{besta2019demystifying}.

\begin{figure}[t]
\centering
\begin{center}
    \includegraphics[width=\columnwidth]{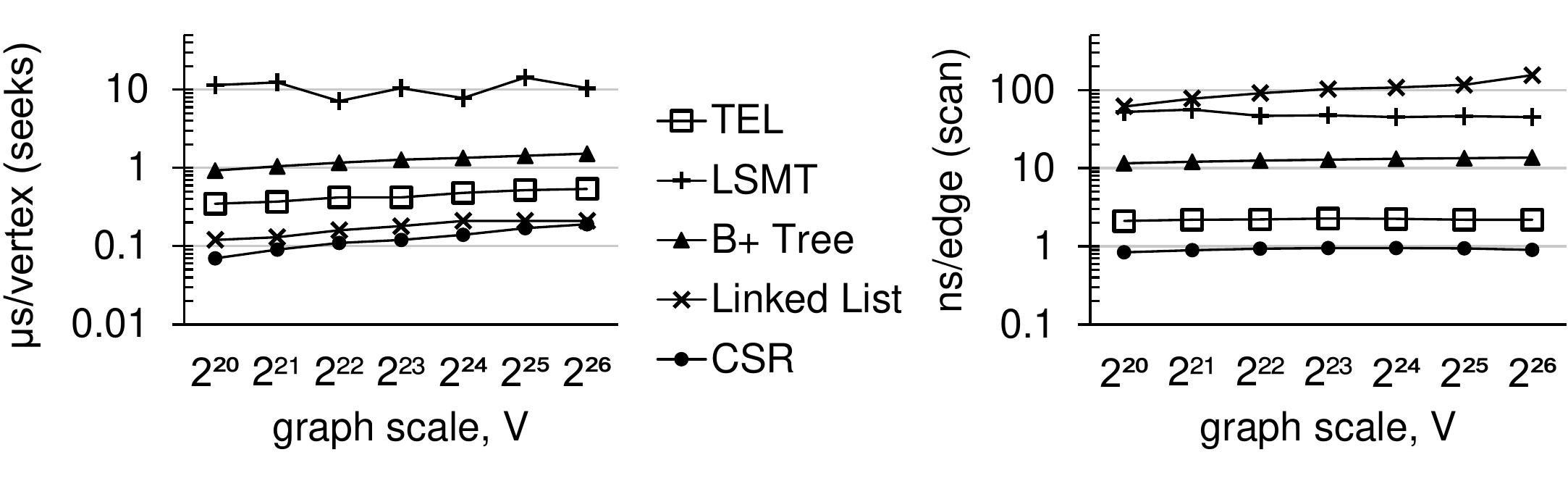}
\end{center}
\begin{subfigure}[b]{0.55\columnwidth}
	\centering
    \caption{Seek latency \label{fig:micro_seek}}
\end{subfigure}
~
\begin{subfigure}[b]{0.4\columnwidth}
	\centering
    \caption{Edge scan latency \label{fig:micro_scan}}
\end{subfigure}
\caption{Adjacency list scan micro-benchmark}
\label{fig:microbenchmarking_results}
\end{figure}

\spara{Micro-benchmark results.}
To see the impact of data structure choices on actual graph systems,  we use a micro-benchmark that performs $10^8$ adjacency list scans, where each start vertex is selected randomly under a power-law distribution. Graphs are generated using the Kronecker generator~\cite{leskovec2010kronecker} with sizes ranging from $2^{20}$ to $2^{26}$ vertices, and an average degree of 4 (similar to those generated by Facebook's LinkBench~\cite{armstrong2013linkbench} benchmark). All graphs fit in the memory of a single socket. For accurate cache monitoring, we perform single-thread experiments, with the workload driver running on a separate socket.

We evaluate LMDB~\cite{lmdb} and RocksDB~\cite{rocksdb}, embedded storage systems that adopt B+ trees and LSMTs respectively.
To fairly isolate the impact of data structure choices, we disable compression in RocksDB, and also implement an efficient in-memory linked list prototype in C++ rather than running Neo4j on a managed language (Java).
For reference, we also include CSR, which is widely used by state-of-the-art graph engines, though CSR is read-only.

We consider two metrics: seek latency and edge scan latency.
Seek latency is the time needed to locate the adjacency list of a vertex.
Edge scan latency is the time required to access the next edge in the adjacency list.
Figure \ref{fig:microbenchmarking_results} shows that using different data structures results in orders of magnitude gaps in these metrics.

The sequential data layout of \LOG is clearly superior to other pointer-based data structures.
To locate the first edge of the range (seek), B+ trees have a logarithmic number of random accesses.
RocksDB's implementation of LSMTs uses a skip list as memtable, which has a similar behavior for seeks.
However, LSMTs also require reading the (on-disk) SST tables for seeks, since only the first part of the edge key is known (the source vertex ID) while the second part is unknown (the destination ID).
This explains the bad performance of LSMT in Figure~\ref{fig:micro_seek}.
For linked-lists and \LOGs we consider one data structure instance per adjacency list, as done in Neo4j and \SYS, respectively. For CSR, the beginning of an adjacency list is stored in the offset array. 
Reading it requires only a constant-time index lookup.

Systems using B+ trees and LSMTs store edges in a single sorted collection, which corresponds to a single edge table.
An adjacency list scan becomes a range query where only the first component of an edge key, the source vertex ID, is given.
To iterate over all edges of a vertex (scans), a B+ tree sequentially scans elements of the same node but it needs random accesses whenever an adjacency list spans multiple nodes.
LSMTs require scanning SST tables also for scans because, similar to seeks, only the first component of the edge key is known.
Skip Lists and Linked List require random accesses for each edge scan.
By contrast, scans in \LOG and CSR are always sequential.
Figure~\ref{fig:micro_scan} shows that \LOG has a scan speedup larger than $29 \times$ over linked list, $20\times$ over LSMT, and $6\times$ over B+ tree. 
CSR's latency for scans is 43\% that of \LOG.
The gap is mainly from \LOG's larger memory footprint and the overheads of checking edge visibility to support transactions (our double timestamps design, which we will discuss later).

We performed a more detailed micro-architectural analysis on the $2^{26}$ scale graph to further understand the behavior of data structures.
B+ trees and LSMTs trigger 7.09$\times$ and 11.18$\times$ more last-level cache misses than \LOG. 
Linked Lists with mostly random memory accesses are the worst, incurring 63.54$\times$ more LLC-misses than \LOG.
\LOG has a simpler sequential execution flow, leveraging CPU pipelining+prefetching and reducing branch mispredictions.
This profiling confirms the huge gap between pointer-based data structures and a sequential data structure like \LOG.
Compared with CSR, \LOG triggers 2.42$\times$ more LLC-misses due to our memory footprints: a single edge in \LOG takes 2$\times$ memory than CSR in our micro-benchmark.


In total adjacency list scan latency, \LOG yields, on average among different graph scales, a $22 \times$ performance improvement over LSMT, $46 \times$ over linked list, $5.6 \times$ over B+ tree, and 40\% higher than CSR.
For seeks, there is a significant gap between tree-based (logarithmic) and graph-aware (constant) data structures. 
For scans, \LOG performs much better as its accesses are purely sequential, while others involve random accesses and additional branching. 

These results show that existing dynamic data structures used by transactional systems leave a lot of performance on the table for graph workloads, which can be harvested by using a graph-aware storage layout like \LOG.




\subsection{Transactions with Sequential Access}

The previous experiments show the importance of eliminating random accesses during adjacency list scans through graph-aware data structures.
However, they consider a single-threaded setting without concurrency.
In transactional workloads, it is necessary to preserve the benefits of sequential scans in the presence of concurrent transactions.

Real-time graph analytics feature read-only transactions that need to access a consistent snapshot of the graph.
These transactions may be long-running and access a large fraction of the graph, but they should not hinder the progress of concurrent transactions.
Multi-versioning is a common approach to ensure this property and many efficient concurrency control algorithms have been proposed for relational databases~\cite{faleiro2015rethinking, larson2011high, lim2017cicada, neumann2015fast, wu2017empirical}. 
However, 
they target data with flat layouts (i.e., indexed records).
Representing each adjacency list as a whole record with multiple versions would make edge updates prohibitively expensive because every time the entire list would have to be written again.

Enabling multi-versioned sequential accesses to adjacency lists is to our knowledge an open topic encompassing multiple challenges: (1) different versions of edges belonging to the same adjacency list must be stored in contiguous memory locations; (2) finding the right version of an edge should not require auxiliary data structures that, in turn, require random access to be visited; (3)  the concurrency control algorithm should not require random access during scans.

\SYS is the first system that guarantees these properties, achieved by co-designing a graph-aware data structure (Section~\ref{sec:representation}) and the concurrency control algorithm (Sections~\ref{sec:ops} and~\ref{sec:transactions}) to ensure purely sequential scans even in the presence of concurrent transactions.

\section{LiveGraph Data Layout}
\label{sec:representation}

\SYS implements both in-memory and out-of-core graph storage on a single server.
It adopts the \textit{property graph model}~\cite{robinson2015graph}, where each object (vertex or edge) can have associated properties (i.e., arbitrary key-value pairs).

Edges have a special type of property called \emph{label}.
Each edge can have only one label.
Edges that are incident to the same vertex are grouped into one adjacency list per label.
Labels can be used to separate edges that are scanned separately, e.g., ``is-friend-with" and ``has-posted" edges in a social graph.
For simplicity, our discussion depicts the case 
where all edges have the same label.

Edge storage is particularly critical since (1) usually graphs have
more edges than vertices and edge operations are more frequent~\cite{bronson2013tao}, and (2) efficient edge scan is crucial, as shown earlier.
Therefore, \SYS adopts a 2-D approach:
vertices are stored individually whereas all the edges incident on the same vertex are grouped in a single Transactional Edge Log (\LOG).


\begin{figure}[t]
\centering
\includegraphics[width=0.6\columnwidth]{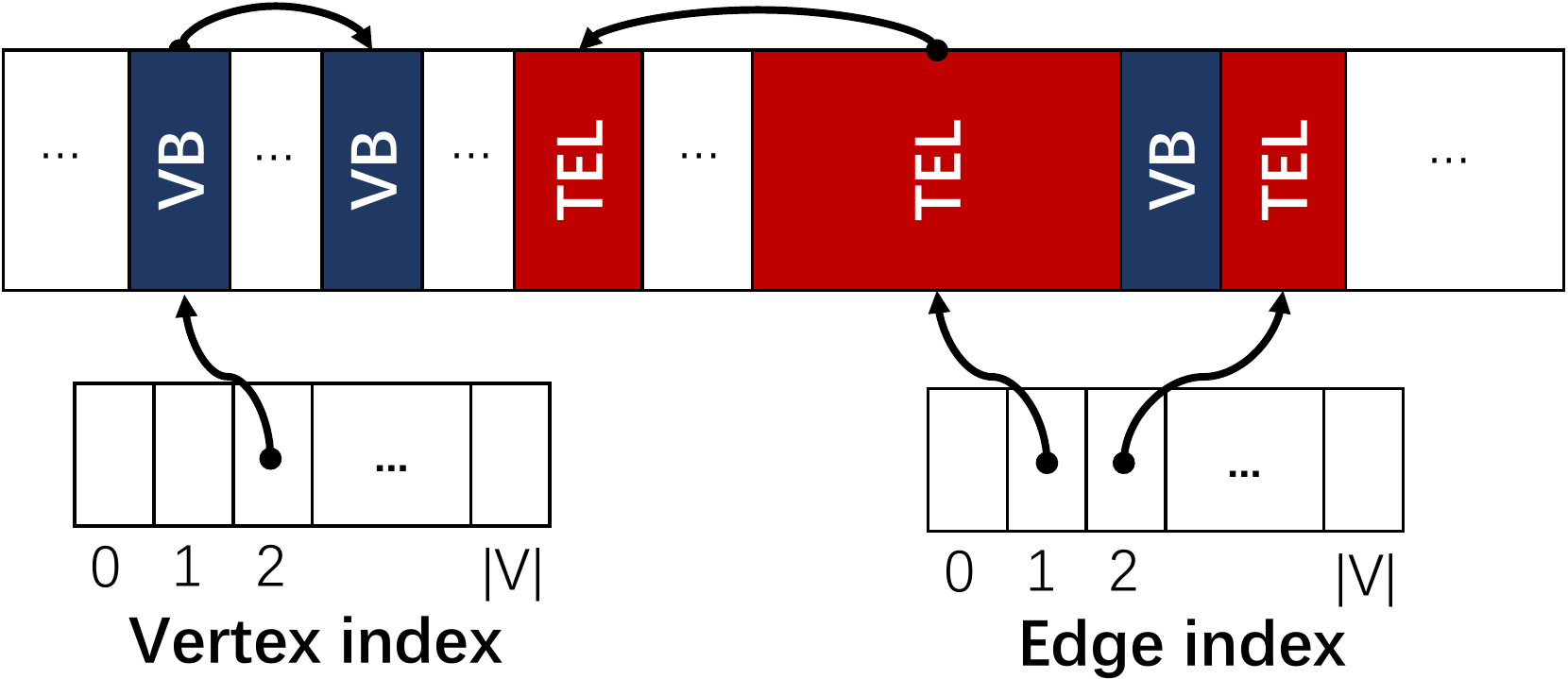}
\vspace{-5pt}
\caption{\SYS data layout. For simplicity, here we omit label index blocks and the vertex lock array.}
\label{fig:data-layout}
\vspace{-10pt}
\end{figure}

Figure~\ref{fig:data-layout} shows the data structures of \SYS, which mainly consist of {\em vertex blocks} (VB) and {\em \LOGs}. 
These are stored in a single large memory-mapped file managed by \SYS's memory allocator.
The blocks are accessed via two index arrays, 
a {\em vertex index} and an \textit{edge index}, storing pointers to appropriate blocks by vertex ID.
Though not depicted in Figure~\ref{fig:data-layout}, there is an additional level of indirection between the edge index and \LOGs, called {\em label index blocks}, used to separate the storage of per-vertex edges with different labels.
Since vertex IDs grow contiguously, we use extendable arrays for these indices.
We have not found this to be a limiting factor in any of our experiments.

\spara{Vertices.}
\SYS stores each vertex separately into the vertex block.
Updates to vertices are relatively infrequent and transactions typically access the latest version.
Therefore, for vertices we use a standard copy-on-write approach, where the newest version of the vertex can be found through the vertex index, and each version points to its previous version in the vertex block.

\spara{Adjacency lists.}
A \LOG is a fixed-size memory block with free space that is resized when filled.
Different versions of a \LOG are linked with ``previous" pointers like vertex blocks.
This organization combines efficient sequential scans of read-optimized formats for analytics (like CSR) with support for updates of dynamic arrays.
Instead of just storing edges constituting the current adjacency list, a \LOG represents all edge insertions, deletions, and updates as log entries appended at the tail of the log.
Note that while our discussion focuses on using \LOG for adjacency list storage, ideas proposed here can be used to implement a general key-value set data structure with sequential snapshot scans and amortized constant-time inserts.

\begin{figure*}[hbt]
\centering
\includegraphics[width=0.95\textwidth]{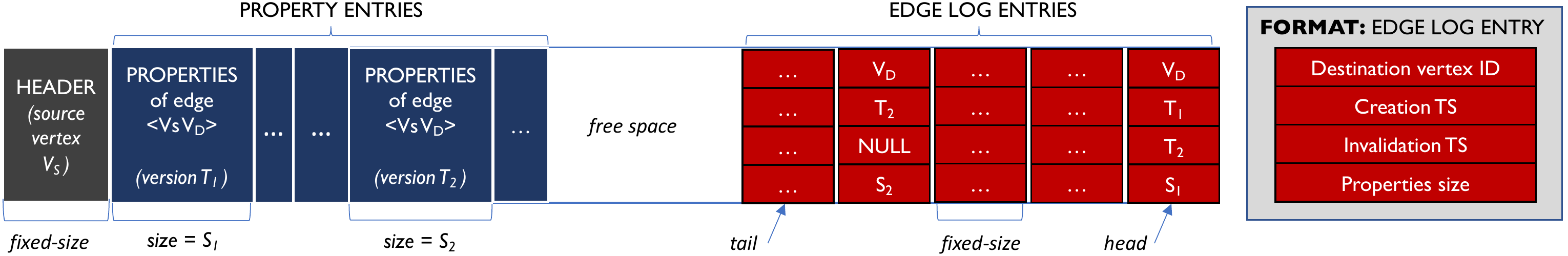}
\caption{A simplified view of a Transactional Edge Log block. A \LOG stores the adjacency list of a specific source vertex, indicated in the \LOG header. Edge log entry fields are cache-aligned, as required by the transaction processing algorithm. In this example, the adjacency list is for vertex $V_S$ and it contains five entries. An edge to vertex $V_D$ is created at time $T_1$ (head entry). The edge is later updated at time $T_2$. The previous entry is invalidated and a new property block is appended.}
\label{fig:graph}
\end{figure*}

The layout of a \LOG block is depicted in Figure~\ref{fig:graph}.
Edge log entries are appended backwards, from right to left, and scanned forwards, from left to right.
This is because many scan operations benefit from time locality, as in Facebook's production workload~\cite{armstrong2013linkbench}, where more recently added elements are read first.
Edge log entries have fixed size with cache-aligned fields.
This is required by our transaction processing protocol, as to be discussed later.
Each entry has two timestamps, a creation timestamp and an invalidation timestamp, indicating its lifecycle.
Edge properties have variable lengths and are appended from the beginning of the block forwards.
These property entries are stored separately from the edge log entries to preserve the data layout alignment of the latter, again as required by transaction processing.
Their content is opaque to \SYS.

For a new vertex, its adjacency list starts small, with 64-byte blocks that accommodate a single edge in our implementation.
When a block is full, \SYS copies the log to an empty block of twice its current size.
Similar to dynamic arrays, appends to the log have a constant amortized cost.
The worst-case cost is linear, but copying contiguous blocks of memory is fast and does not result in high tail latency, as our evaluation shows.

This design is particularly suited to 
power-law degree distributions that are common in real-world graphs.
The majority of vertices, being low-degree and less active, will grow slowly in degree, with limited space waste.
The high-degree vertices are likely to grow faster and incur higher data copying costs when ``upgraded'', but such relocation happens at decreasing frequency with exponentially growing block sizes.
The power-law degree distribution also implies that only a small fraction of vertices occupy large log blocks.
We describe the details of \SYS's memory management and log relocation in Section~\ref{sec:storage}.


\section{Single-Threaded Operations}
\label{sec:ops}
\label{sec:basic}

We first describe \SYS's operations in the absence of concurrency.
\SYS uses a multi-versioned data store and each transaction is associated with a read timestamp, which determines the version of the data it operates on.
This timestamp is determined by the transaction processing mechanism (to described in Section~\ref{sec:transactions}).

\spara{Vertex operations.}
Vertex reads access the vertex index and then the vertex block; writes create a new version of the vertex block in the block storage, including a pointer to the current version, and set the vertex index to point to the new version.
In the uncommon case where a read requires a previous version of the vertex, it follows the per-vertex linked list of vertex block versions in backward timestamp order until it finds a version with a timestamp smaller than its read timestamp.

Adding a new vertex first uses an atomic \texttt{fetch-and-add} operation to get the vertex ID, fills in empty pointers in the corresponding locations of vertex and edge indices, and sets the lock status.
If the vertex needs to store properties or add edges, it asks the storage manager to allocate a block according to the size it needs, whose details will be shown in Section~\ref{sec:storage}.
Garbage collection to reuse the IDs of deleted vertices can be achieved by using techniques described later in Section~\ref{sec:storage}.
Since vertex deletions are rare, we leave the implementation of this mechanism
to future work.

\spara{Sequential adjacency list scans.}
Scanning adjacency lists efficiently is a key requirement of analytics workloads.
\SYS achieves purely sequential adjacency list scans by combining log-structured storage of the adjacency list and double timestamps.

A log structure is a more convenient multi-versioned representation of the adjacency list compared to alternative approaches.
For example, a coarse-grained copy-on-write approach to multi-versioning would create a new copy of the entire adjacency list \emph{every} time an edge is updated.
This is the approach used by Grace~\cite{prabhakaran2012managing}.
However, it makes updates very expensive, especially for high-degree vertices.

Storing multiple versions of the adjacency list in contiguous memory locations as a log is key to achieving purely sequential adjacency list scans, but it is not sufficient.
The same edge
can now correspond to multiple entries in the log, as shown in Figure~\ref{fig:graph}.
When a thread executing an adjacency list scan reads an edge entry, it cannot tell whether the entry is still valid or if it has been deleted or modified at a later position in the log.
The thread could keep a hash map to progressively update the latest version of each edge during the scan.
But accessing the map would again require random accesses, which we strive to avoid.

To this end, \SYS stores a double timestamp for each edge.
A read operation with timestamp $T$ considers only edge entries such that $T$ is within the entry's creation and invalidation timestamps.
These two timestamps determine if the entry is valid for the read timestamp of the transaction (see Figure~\ref{fig:graph}).
Such a design makes \LOG scans sequential: a transaction can check the validity of an edge entry simply by checking the embedded timestamp information.
Scans that access edge properties require two sequential scans, one forwards from ``tail'' (as shown in Figure~\ref{fig:graph}) to the right end, for edge log entries, and one backwards from the end of the property entries, for properties.

\spara{Edge updates and constant-time insertions.}
To support fast ingestion of new edges, \SYS inserts edges in amortized constant time.
Insertions in a regular log-based data structure are simple appends, which can be done in constant time.
\LOGs sometimes require resizing in order to append new entries, but the amortized cost of appends is still constant, like in dynamic arrays (see Section~\ref{sec:representation}).
The dual timestamp scheme of \SYS, while useful for sequential scans, makes updating the adjacency list more complex.
Appending log entries is not always sufficient any longer.
If an operation updates or deletes an existing edge, it must also update the invalidation timestamp of the previous entry for that edge, which entails scanning the log.
However, if a new edge is inserted, the scan is not necessary and a constant-time append is still sufficient.

\SYS includes a Bloom filter in the \LOG header to determine whether an edge operation is a simple insert or a more expensive update.
Inserting a new edge appends an entry at the end of a \LOG and updates the Bloom filter as well as the adjacency list block size.
Edge deletions/updates first append a new entry to the \LOG and check, using the Bloom filter, if a previous version of the edge is present in the \LOG.
If so, its invalidation timestamp needs to be updated.
The Bloom filter is also handy to support fast ``upsert'' semantics (such as in the Facebook LinkBench workload~\cite{armstrong2013linkbench}), where a lookup is needed to check whether the entity to be inserted already exists.
It allows distinguishing which of them are ``true insertions'' that add new edges, such as ``likes'' in social networks or new purchases in online stores.
Inserts often compose the majority of write operations and \SYS processes them in amortized constant time.

Upon a ``possibly yes'' answer from the Bloom filter, however,
finding the edge itself involves a tail-to-head \LOG scan, traversing the entire adjacency list in the worst case.
In practice though, edge updates and deletions have high time locality: edges appended most recently are most likely to be accessed.
They are the closest to the tail of the \LOG, making average cost fairly low for edge updates/deletions.


Each Bloom filter is fixed-sized: 1/16 of the \LOG for each block larger than 256 bytes.
A blocked implementation~\cite{putze2007cache} is adopted for cache efficiency considerations.

\spara{Reading a single edge.}
Reading a single edge involves checking if the edge is present using the Bloom filter.
If so, the edge is located with a scan.
As with adjacency list scans, the scan skips any entries having a creation-invalidation interval inconsistent with the read timestamp.

%

\section{Transaction Processing}
\label{sec:transactions}
Next, we discuss concurrent execution of transactions, each consisting of one or more of the \emph{basic read/write operations} described in Section~\ref{sec:ops}. 
The major innovation of \SYS here lies in its integrated design: 
unlike traditional MVCC solutions~\cite{wu2017empirical}, 
which adopt auxiliary data structures to implement concurrency control, 
\SYS embeds the information required for transactional adjacency list scans within its main \LOG data structure.
This leads to
significant performance advantages because it enables  sequential access even in the presence of concurrent transactions.

We first present this integrated data layout, followed by the corresponding transaction management algorithms.  

\spara{Data layout and coordination.}
\SYS stores adjacency lists as multi-versioned logs to efficiently support \textit{snapshot isolation}~\cite{berenson1995critique}, which allows read operations to proceed without interfering with each other and with write operations.
Snapshot isolation is stronger than \textit{read committed}, the default isolation level of Neo4j and many other commercial DBMSs~\cite{bailis2013highly}.

In \SYS, basic read operations on edges do not acquire locks.
Coordination with basic write operations on edges occurs only through \emph{cache-aligned 64-bit word timestamps}, written and read atomically.
Cache-alignment is done by separating the variable-size edge properties and fixed-size edge log entries, which grow from the opposite ends of \LOG blocks (Figure~\ref{fig:graph}).
Thus all timestamps in the log entries are cache-aligned.
Basic read operations on vertices access different versions as described in Section~\ref{sec:ops}.

Two additional cache-aligned per-\LOG variables are used for coordination: the log commit timestamp \texttt{CT} and the log size \texttt{LS}.
They are both stored in a \LOG's fixed-size header.

Write-write conflicts are detected using per-vertex locks, implemented with a \texttt{futex} array of fixed-size entries (with a very large size pre-allocated via \texttt{mmap}).
We also explored other choices, such as concurrent hashtables or spinlock arrays, but found the futex array method most scalable.
For write-intensive scenarios when many concurrent writers compete for a common lock, spinning becomes a significant bottleneck while futex-based implementations utilize CPU cycles better by putting waiters to sleep.
The space consumption for locks is acceptable: under $0.5\%$ of the overall dataset size in our evaluation.

\begin{figure*}[tp]
\centering
\begin{subfigure}[b]{\columnwidth}
	\centering
    \includegraphics[width=0.95\columnwidth]{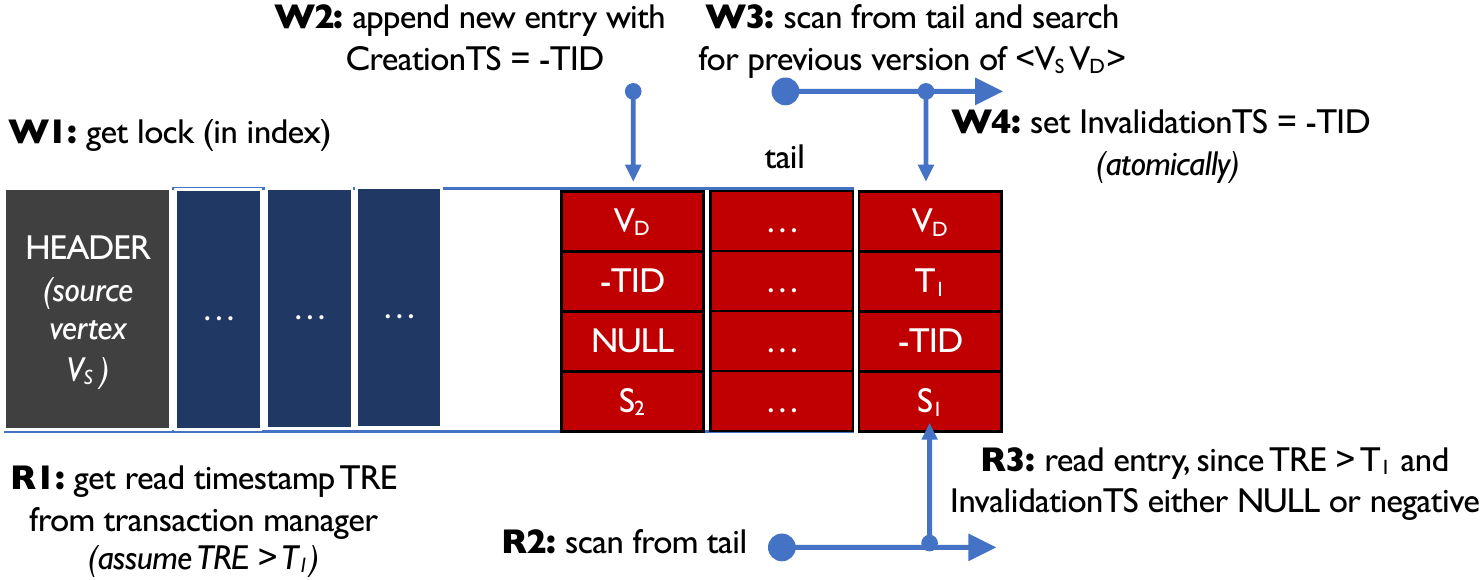}
    \caption{Work phase \label{fig:working}}
\end{subfigure}
~
\begin{subfigure}[b]{\columnwidth}
	\centering
    \includegraphics[width=0.95\columnwidth]{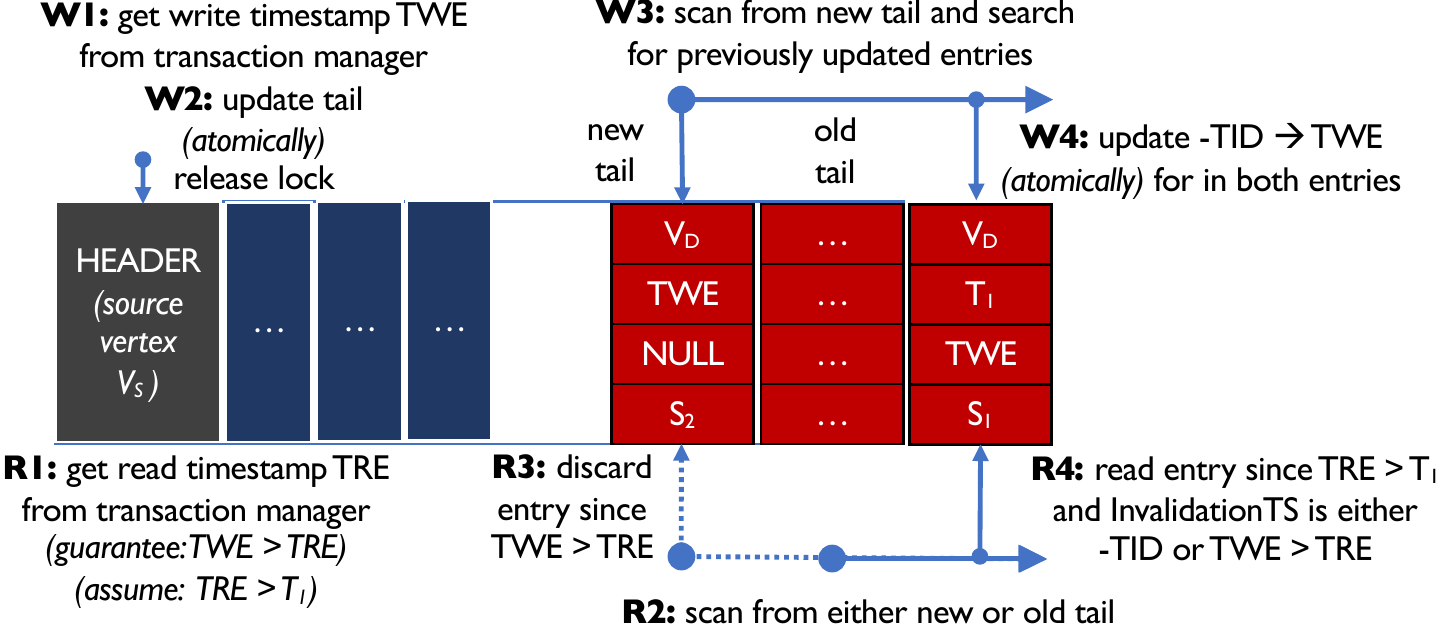}
    \caption{Apply phase \label{fig:apply}}
\end{subfigure}
\caption{Example of concurrent access to the same \LOG. A transaction is executing a basic write operation (see notes above the \LOG) while another concurrent transaction executes a basic read operation (notes below) on the same \LOG. 
Each operation is illustrated in steps.
Concurrency can happen in the work/apply phase of the writing transaction. 
Accesses to cache-aligned timestamps are guaranteed to be atomic.}
\label{fig:example}
\end{figure*}

\spara{Transactions.}
\SYS keeps a pool of transaction-serving threads (henceforth referred to as ``workers''), plus one transaction manager thread.
All threads share two global epoch counters, \texttt{GRE} for reads and \texttt{GWE} for writes, both initially set to $0$.
They also share a reading epoch table to establish a safe timestamp for compaction.

A transaction has two local variables: a transaction-local read epoch counter \texttt{TRE}, which the worker initializes to \texttt{GRE}, and a transaction-local write epoch counter \texttt{TWE}, which is determined by the transaction manager at commit time.
The worker assigns to the transaction a unique transaction identifier \texttt{TID}, by concatenating its unique thread ID and a worker-local logical transaction count.

Basic read operations access a snapshot determined by their \texttt{TRE}.
When they read a vertex, transactions visit the linked list of vertex versions until they find the right version.
In practice, a transaction needs to visit the list only in the rare case in which there are concurrent vertex writes.
Reads need not acquire locks. 

Adjacency list scans are strictly sequential. 
When they scan a \LOG, read operations only consider edge log entries such that either \texttt{(0 <= CreationTS <= TRE) AND ((TRE < InvalidationTS) OR (InvalidationTS < 0))} (entry is valid at time \texttt{TRE}), or \texttt{(CreationTS == -TID) AND (InvalidationTS != -TID)}
(a transaction sees its own earlier writes, more details below).
The scan starts from the end of the log, using the \texttt{LS} size variable within the \LOG header, and proceeds backwards towards the beginning.

Basic write operations initially store their updates in a transaction-private working version, 
in the same \LOG as committed updates, to keep adjacency list scans sequential.
Transaction-private updates are
made visible to other transactions only after commit.

A write transaction goes through three phases: \emph{work}, \emph{persist}, and \emph{apply}.
In the work phase, it starts by computing its unique transaction identifier \texttt{TID} and acquiring write locks. To avoid deadlocks, a simple timeout mechanism is used: a timed-out transaction has to rollback and restart.
The write transaction then executes write operations as discussed in Section~\ref{sec:basic}, using the timestamp \texttt{-TID} to make the updates private.
This phase concludes by calling the transaction manager to persist the changes.

In the persist phase, \SYS adopts standard group commit processing~\cite{DBLP:books/aw/BernsteinHG87} to enhance throughput. 
The transaction manager first advances the \texttt{GWE} counter by 1, then appends a batch of log entries to a sequential write-ahead log (WAL) and uses \texttt{fsync} to persist it to stable storage.
Next, it notifies the involved transactions by assigning them a write timestamp \texttt{TWE}, set as \texttt{GWE}.

Now the write transaction enters the final apply phase, 
by updating the commit timestamp of the \LOG, which is set to \texttt{TWE}, and the log size \texttt{LS} in the header.
For vertex blocks, it also updates the version pointers and sets the vertex index pointer to the latest version.
Next, it releases all its locks before starting the potentially lengthy process of making its updates visible by converting their timestamps from \texttt{-TID} to \texttt{TWE}.
After all transactions in the commit group make their updates visible, the transaction manager advances the global read timestamp \texttt{GRE}, exposing the new updates to upcoming transactions.
This also guarantees that the read timestamp of a transaction is always smaller than the write timestamp of any ongoing transaction.

Figure~\ref{fig:example} shows an example of a write operation executed concurrently with reads in two of its phases (the persist phase is done by the transaction manager/logger and omitted due to space limitations).
It shows that (1) read operations do not need to use locks and (2) multi-version transaction management is done as part of the sequential \LOG scan. Both aspects significantly contribute to \SYS's processing speed and throughput.

To ensure consistent versioning, it is necessary to guarantee the invariant that each entry's invalidation timestamp is always larger than its creation timestamp.
Write operations having a timestamp $T$ must abort if they try to invalidate an entry with creation timestamp larger than $T$.
\SYS stores the latest timestamp that has modified a \LOG in a commit timestamp variable \texttt{CT} in the \LOG header.
This way, write operations can simply compare their timestamp against \texttt{CT} instead of paying the cost of scanning the \LOG only to later find that they must abort.


Whenever a transaction aborts, it reverts the updated invalidation timestamps from \texttt{-TID} to \texttt{NULL}, releases all acquired locks, and returns any newly allocated blocks to the memory manager.
An aborted transaction never modifies the log size variable \texttt{LS} so its new entries will be ignored by future reads and overwritten by future writes.

\spara{Correctness.}
We now argue that \SYS guarantees snapshot isolation~\cite{berenson1995critique}.
Snapshot isolation rules out the following anomalies: dirty writes, dirty reads, read skew, and phantom reads.
In the following, we show that these anomalies cannot occur.
We consider vertices and adjacency lists as the atomic variables over which the consistency properties must hold.

A \textit{dirty write} occurs when a transaction updates a variable and another transaction modifies it before the former commits.
This is avoided by having basic write operations acquire locks before modifying vertices or adjacency lists, and release it only after commit.

A \textit{dirty read} occurs if a transaction reads a value written by another transaction that has not yet committed.
For vertices, dirty reads are avoided because write operations make a new vertex version, reachable through the vertex index only after committing.
Each vertex pointer is a cache-aligned word so its updates are atomic: concurrent reads either see the old version or the new one.
For adjacency lists, if a read has a timestamp \texttt{TRE} greater than the one of the write transaction, then the write transaction has committed and completed updating the adjacency list in the apply phase, therefore the read is not dirty. 
Otherwise, the read timestamp is lower than the write timestamp, in which case the read simply ignores the entries updated by the write transaction, as illustrated by Figure~\ref{fig:example}. 
This is because writes append past the tail pointer and update the tail pointer atomically after commits, so the read transaction never reads an incomplete edge log entry.
Similarly, uncommitted updates to adjacency list entries initially have a negative creation/invalidation timestamp.
A write transaction sets these timestamps to the write timestamp, which is larger than \texttt{TRE}, only after it commits.
Both timestamps are cache-aligned words, guaranteed to be read/updated atomically.

\textit{Read skew} occurs when a transaction $A$ reads a variable $x$, then another transaction $B$ writes to $x$ and $y$ and commits, after which $A$ reads the value of $y$ written by $B$.
We now show that with \SYS design, $A$ in this case cannot not read the value of $y$ written by $B$.
The value of the read timestamp \texttt{TRE} for $A$, $r_A$, is set to the value of the global read counter \texttt{GRE} when $A$ starts.
The global write counter \texttt{GWE} is always greater than or equal to \texttt{GRE}, therefore $r_A\!\leq$\texttt{GWE}.
Transaction $B$ writes values after $A$ reads $x$, so $B$ was not committed when $A$ started.
When $B$ later commits, it is assigned a write timestamp $w_B$, which is greater than the value of \texttt{GWE} when $A$ started, therefore we have $w_B\!>\!r_A$.
Read skew cannot occur because the condition for $A$ to consider the value a variable written by $B$ is not satisfied, 
following an argument analogous to the one above for dirty reads.
If $y$ is a vertex, $A$ ignores the version written by $B$ since $r_A\!<\!w_B$ and moves to the next version in the linked list.
If $y$ is an adjacency list, the edge entries written by $B$ have either a negative creation timestamp or a creation timestamp $w_B > r_A$, so they are ignored by $A$.
Entry invalidations are also ignored for similar reasons.

\textit{Phantom reads} occur when a transaction $A$ reads a set of variables that satisfy a predicate, then a transaction $B$ adds or deletes a variable that satisfies that predicate and commits, and finally transaction $A$ issues again a read on the same predicate and obtains a different set of variables.
This cannot happen based on an argument similar to read skew, as it is the special case where $B$'s write to $y$ is an insertion/deletion and $A$'s read is based on a predicate.

\section{Storage Management}
\label{sec:storage}


Currently, \SYS uses a memory-mapped file to store graph data, comprising vertex blocks, which store single vertices, and \LOG blocks, each storing a \LOG (see Figure~\ref{fig:data-layout}), 
as well as per-vertex label index blocks storing pointers to \LOGs with different labels of each vertex.
This allows \SYS to store graphs that do not fit in memory and rely on the operating system to decide how data is cached in memory and evicted.
We plan to replace \texttt{mmap} with a managed page cache~\cite{leis2018leanstore} to enable more robust performance on very large datasets backed by high-speed I/O devices.

\spara{Compaction.}
The \LOG is not just an adjacency list but also, implicitly, a multi-version log record of the adjacency list with entries sorted by creation timestamps.
Invalidated entries are useful for retrieving and analyzing historical snapshots, but their accumulation will eventually bloat the \LOG size and impede in-memory processing.
Therefore, \SYS performs periodic compaction.

\SYS provides the capability of a user-specified level of historical data storage, trading off disk space and checkpointing overhead, to allow full or partial historical snapshot analysis.
In the prototype evaluated, we performed rather aggressive garbage collection (GC), without saving invalidated copies to disk.
\SYS periodically (every 65536 transactions in our default setting) launches a compaction task.
Each worker thread in \SYS maintains a \emph{dirty vertex set}, marking vertices whose corresponding blocks have been updated since the last compaction executed within this thread.
When doing compaction, a thread scans through its local dirty set and compacts or garbage-collects blocks based on \emph{version visibility}.
Each worker stores the transaction-local read epoch counter (the \texttt{TRE}) used by its \emph{ongoing} transaction (if any) in a table with one entry per worker.
\emph{Future} transactions will get a \texttt{TRE} that is greater or equal to the current global read epoch counter (the \texttt{GRE}).
A thread doing compaction accesses all these epoch counters to determine version visibility for all transactions.

The compaction processes one block at a time, asynchronously and independently, with only minimal interference with the regular workload.
If a \LOG block is not visible any longer to any ongoing or future transaction, it is garbage-collected.
Otherwise, the thread removes all entries that will not be visible to future transactions.
A thread first scans the block to compute the new capacity (sometimes the block could shrink after many edges being deleted).
A new block is then allocated and only the entries visible to future transactions are copied (sequentially) to this new block.
Like in a regular write transaction, the corresponding vertex lock of the \LOG is held to temporarily prevent concurrent writes to that specific block.
Writes to the new blocks are also committed like with write transactions.
Once the copy to the new \LOG block is committed, the compacting thread releases the lock and moves to the next vertex in the dirty set.
New transactions that start after the commit access the new version.  Ongoing transactions continue having read-only access to the old versions, which are kept until they are no longer visible by any transaction.
At that point, in a future compation cycle, the thread garbage-collects them.
Compaction only occurs in a lightweight vertex-wise fashion: unlike LSMT, \SYS never needs to compact multiple on-disk files through merging.

Compactions for vertex blocks are similar to existing MVCC implementations.
Invalidated blocks that will never be visible to any existing or future transactions are simply garbage collected (to be reused later) and any related ``previous pointers" are cleared simultaneously.

%
%

\spara{Space overhead for timestamps.}
Using two timestamps (which are not explicitly used by graph analytics algorithms themselves) dilutes \LOG's storage density and lowers its bandwidth/caching efficiency compared to compact data structures such as CSR.
This results in a performance gap between running analytics on top of \SYS compared to state-of-the-art engines for static graphs such as Ligra~\cite{shun2013ligra} or Gemini~\cite{zhu2016gemini}.
However, analytics on top of \SYS do not need to perform expensive ETL (Extract-Transform-Load) operations to load the graph into a dedicated tool.
Compared to systems that support transactional graph updates and use pointer-based data structures, \SYS uses sequential memory regions and thus saves the storage cost of keeping many pointers, in addition to supporting multi-versioning.
Overall, our evaluation shows that \SYS has a similar memory overhead as these systems.

\spara{Memory management.}
In selecting the adjacency list block size, there is a trade-off between the cost of repeated data relocation and space utilization. 
In making this decision, we seize a rare opportunity offered by the power-law degree distribution found in many real-world graphs~\cite{chen2015powerlyra,faloutsos1999power,soramaki2007topology}.
Inspired by the buddy system~\cite{knowlton1965fast}, \SYS fits each \LOG into a log block of the closest power-of-2 size. 
 


\SYS has \LOGs starting from a size of 64 bytes (a 36-byte header plus a 28-byte log entry, whose contents were described earlier).
This minimal configuration accommodates one edge
and occupies one cache line in common processors today.
An array of lists \texttt{L} is used for keeping track of the free blocks in the block store, where \texttt{L[i]} ($i=0, 1, ..., 57$) contains the positions of blocks with size equal to $2^i\times64$ bytes.
When a block of a certain size is needed, \SYS first checks
the corresponding free list, allocating new blocks from the tail of the block store only when that list is empty.
Vacated blocks or those that do not contain any valid data, meanwhile, are recycled into the proper free lists.

Again considering the power-law degree distribution,
we accelerate the allocation process by differentiating the free list management of smaller and larger blocks.
We define a tunable threshold $m$, with each thread maintaining its private free list array \{\texttt{S[0], ..., S[m]}\} and sharing a global free list array
\{\texttt{S[m+1], ..., S[57]}\}.
This significantly reduces the contention over the free lists for allocating highly popular small blocks, while mitigating waste by centralized large block management.
Measurements on our 24-core (48 hardware threads) test platform show that block allocation is not a performance bottleneck (with $m$ set at 14).

\spara{Recovery.} The recovery mechanism of \SYS is similar to write-ahead-logging systems. A checkpointer (which can be configured to use any number of threads) periodically persists the latest consistent snapshot (using a read-only transaction) and prunes the WAL entries written before the epoch of the snapshot. When a failure happens, \SYS first loads the latest checkpoint and then replays the WAL to apply committed updates.

\section{Evaluation}
\label{sec:eval}
\subsection{Experimental Setup}
\label{subsec:setup}
\para{Platform.}
We set up experiments on a dual-socket server, whose specification is given in Table~\ref{tab:testbed_spec}. The persistence features are enabled for all the systems, except when specified otherwise.
We evaluate two alternative SSD devices for persistence, with the Intel Optane as the default.

\begin{table}[!hbt]
\center
\caption{Testbed specification}
\label{tab:testbed_spec}
\begin{tabularx}{0.48\textwidth}{c|Y}
\hline
\multirow{2}{*}{Processor} & 2-socket Intel Xeon Gold 6126 CPU \\
& 12 cores per CPU  \\
\hline
Memory & 192 GB DDR4 RAM \\
\hline
\multirow{2}{*}{Storage} & Intel Optane P4800X 750 GB SSD \\
 & Dell Express PM1725a 1.6 TB SSD \\
\hline
\end{tabularx}
\end{table}



\para{Workloads.}
For lightweight graph accesses, we use LinkBench~\cite{armstrong2013linkbench}, a Facebook benchmark based on its social graph interactive workloads.
Besides its default configuration (\textit{DFLT}) with 69\% read and 31\% write operations, we add a ``read-mostly'' workload (\textit{TAO}) with 99.8\% of reads, with parameters set according to the Facebook TAO paper~\cite{bronson2013tao}.
All experiments start on a 32M-vertex, 140M-edge base graph generated by the LinkBench driver, each client sends 500K query requests to the server.

For real-time graph analytics, we use the interactive workload in LDBC Social Network Benchmark (SNB)~\cite{erling2015ldbc}, which simulates the users' activities in a social network for a period of time.
Its schema has 11 entities connected by 20 relations, with attributes of different types and values, providing a rich benchmark dataset.
The SNB data generator is designed to produce directed labeled graphs that mimic the characteristics of real-world social graphs.
We set 10 as the Scale Factor, with 30M vertices and 177M edges in the generated initial graph.
Its requests are classified into three categories: short reads (similar to LinkBench operations), transactional updates (possibly involving multiple objects), and complex reads (multi-hop traversals, shortest paths, and analytical processing such as filters, aggregations, and joins).
Finally, we run two popular iterative analytical algorithms on top of the generated graph, PageRank and Connected Components (ConnComp).
PageRank runs for 20 iterations, while ConnComp runs till convergence.


\para{Choice and rationale of baselines.}
For LinkBench, we first tested MySQL (v5.7.25) and MyRocks (v5.6.35) using their official adaptor, but found that inter-process communication between the server and client (benchmark driver) amplifies latencies.
Thus we compare \SYS with three \emph{embedded} implementations,\footnote{We record traces collected from MySQL runs and replay them for each system. Thinking times (i.e., the time to generate each request) are also recorded and reproduced.} LMDB (v0.9.22), RocksDB (v5.10.3), and Neo4j (v3.5.4), as representatives for using B+ tree, LSMT, and linked list respectively.
This way we focus on comparing the impact of data structure choices.
For Neo4j, we use its Core API rather than Cypher, to eliminate potential query language overhead.



For SNB, besides graph databases including Neo4j~\cite{neo4j}, 
DBMS S, and DBMS T (anonymized due to licensing restrictions), we also compared with PostgreSQL (v10.7)~\cite{postgresql} and Virtuoso (v7)\footnote{The feature/analytics branch from \url{https://github.com/v7fasttrack/virtuoso-opensource.git}, which is about 10$\times$ faster than v7.2.5 from master branch.}~\cite{virtuoso}, two relational databases.
DBMS S is based on the RDF model and uses a copy-on-write B+ tree similar to LMDB as the storage backend;
DBMS T is a commercial graph database that provides the highest performance among graph databases according to 
benchmark reports;
PostgreSQL is among the most popular relational databases for OLTP;
Virtuoso is a multi-modal database that has published its SNB results (and 
offers state-of-the-art SNB results,
based on our survey and experiments).

The implementations for these systems are included in the official SNB repository~\cite{ldbc_snb_implementations},
except for DBMS T, whose implementation is from its own repository, and
currently only implements read-only queries and the driver is limited to running one type of query each time rather than the mix of concurrent queries spawned by the official driver. Therefore, its throughput is estimated by a weighted sum according to each query's frequency (given by the official driver) and measured average latency.
Neo4j and DBMS S's results are omitted as they are several orders of magnitude slower, echoing the findings in existing literature~\cite{pacaci2017we}.


PostgreSQL and Virtuoso implement the requests with SQL plus stored procedures, and interact with the benchmark driver through JDBC.
DBMS T calls RESTful APIs to serve queries by pre-installed
stored procedures.
\SYS's implementation uses a fixed query plan for each type of request.
We build a server with the Apache Thrift RPC framework to communicate with the benchmark driver.

We enable transactional support on all systems.
\SYS guarantees snapshot isolation.
LMDB and DBMS T do not support concurrent write transactions, so they
provide serializable isolation. We configure the other systems to use either snapshot isolation like \SYS, if available, or a default (usually weaker) isolation level. 
More precisely we use read committed for Neo4j, MyRocks, and PostgreSQL, and repeatable read for MySQL and Virtuoso.

\subsection{Transactional Workloads}
\label{subsec:linkbench-perf}
\begin{table}[t]
\center
\caption{Latency w.\ LinkBench TAO in memory (ms)}
\label{tab:tao_latencies_im}
\resizebox{\columnwidth}{!}{
\begin{tabular}{c|ccc|ccc}
\hline
Storage & \multicolumn{3}{c|}{Optane SSD}             & \multicolumn{3}{c}{NAND SSD}                   \\ \hline
System  & \SYS            & RocksDB & LMDB            & \SYS             & RocksDB  & LMDB             \\ \hline
mean    & \textbf{0.0044} & 0.0328  & \textbf{0.0109} & \textbf{0.0051}  & 0.0309   & \textbf{0.0098}  \\
P99     & \textbf{0.0053} & 0.0553  & \textbf{0.0162} & \textbf{0.0058}  & 0.0526   & \textbf{0.0161}  \\
P999    & \textbf{1.0846} & 4.8716  & \textbf{2.0703} & \textbf{1.1224}  & 4.1968   & \textbf{1.5769}  \\
\hline
\end{tabular}
}
\bigskip
\caption{Latency w.\ LinkBench DFLT in memory (ms)}
\label{tab:linkbench_latencies_im}
\resizebox{\columnwidth}{!}{
\begin{tabular}{c|ccc|ccc}
\hline
Storage & \multicolumn{3}{c|}{Optane SSD}            & \multicolumn{3}{c}{NAND SSD}          \\ \hline
System  & \SYS            & RocksDB         & LMDB   & \SYS    & RocksDB           & LMDB    \\ \hline
mean    & \textbf{0.0449} & \textbf{0.1278} & 1.6030 & \textbf{0.0588}  & \textbf{0.1459}   & 1.6743  \\
P99     & \textbf{0.2598} & \textbf{0.6423} & 9.3293 & \textbf{0.2838}  & \textbf{0.8119}   & 9.8334  \\
P999    & \textbf{0.9800} & \textbf{3.5190} & 12.275 & \textbf{1.4642}  & \textbf{4.8753}   & 13.365  \\ \hline

\end{tabular}
}
\end{table}


\para{In-memory latency.} First we evaluate transaction processing.
Tables~\ref{tab:tao_latencies_im} and \ref{tab:linkbench_latencies_im} give the average latency measured from LMDB, RocksDB, and \SYS in memory, with the LinkBench TAO and DFLT workloads respectively, using 24 client threads for request generation and Optane/NAND SSD for transactional durability.
The average latencies for MySQL/MyRocks/Neo4j using Optane SSD as the backing storage are 0.187/0.214/0.236 ms for TAO, and 0.708/0.280/1.231 ms for DFLT.
As these systems are clearly slower than \SYS, RocksDB, and LMDB, their detailed results and analysis are omitted.

The results demonstrate \SYS's significant performance advantage for both workloads.
For the almost read-only TAO, \SYS improves the average latency by 2.47$\times$ from the runner-up (LMDB).
\LOG's major advantage here comes from storing edges by time order, facilitating fast-backward partial scans returning \emph{latest edges}.
This type of query is common not only in social network workloads~\cite{bronson2013tao} but also in other transactional graph applications (such as traffic maps and financial records).
Not only does this accelerate memory accesses, it also improves both spatial and temporal locality, achieving more effective prefetching.
In addition, compared to B+ trees and LSMTs, \LOG has lower complexity for most operations, and avoids pointer chasing through the use of sequential data structures.

For DFLT, which contains 31\% writes, \SYS remains a consistent winner across all categories.
Due to its write-friendly sequential storage, its margin of advantage is even higher, beating the runner-ups in average, p99, and p999 latency by 2.67$\times$, 3.06$\times$ and 4.99$\times$ respectively.
Here LMDB suffers due to B+ tree's higher insert complexity and its single-threaded writes, while both \SYS and RocksDB benefit from their write-friendly log-structured design.
However, as DFLT's majority of transactions are still reads,  RocksDB's overall performance is severely dragged down by its inferior read performance in-memory.

\SYS has linear complexity (in terms of adjacency list size) when searching for a single edge, as opposed to the logarithmic cost of B+ trees or LSMTs.
However, these operations (i.e., read/update/delete a specific edge of a high-degree vertex) are rare in the two LinkBench workloads.
In particular, insertions can usually (in more than 99.9\% of the cases, as found by our profiling) skip such searches, thanks to early rejection enabled by its embedded Bloom filters.
Therefore, these operations do not impact tail latency much.
\SYS's use of compaction also does not result in significantly higher tail latency than other systems.
This is because its compaction only scans a small subset of blocks: the dirty vertex set maintained by each thread.

\begin{table}[t]
\center
\caption{Latency w.\ LinkBench TAO out of core (ms)}
\label{tab:tao_latencies_ooc}
\resizebox{\columnwidth}{!}{
\begin{tabular}{c|ccc|ccc}
\hline
Storage & \multicolumn{3}{c|}{Optane SSD}             & \multicolumn{3}{c}{NAND SSD}               \\ \hline
System  & \SYS            & RocksDB & LMDB            & \SYS            & RocksDB         & LMDB   \\ \hline
mean    & \textbf{0.0166} & 0.0420  & \textbf{0.0364} & \textbf{0.0725} & \textbf{0.1065} & 0.1322 \\
P99     & \textbf{0.1856} & 0.1135  & \textbf{0.3701} & \textbf{0.4830} & \textbf{0.2535} & 0.8372 \\
P999    & \textbf{1.9561} & 4.9366  & \textbf{3.3600} & \textbf{2.5112} & \textbf{4.6701} & 4.9119 \\ \hline
\end{tabular}
}
\bigskip
\caption{Latency w.\ LinkBench DFLT out of core (ms)}
\label{tab:linkbench_latencies_ooc}
\resizebox{\columnwidth}{!}{
\begin{tabular}{c|ccc|ccc}
\hline
Storage & \multicolumn{3}{c|}{Optane SSD}             & \multicolumn{3}{c}{NAND SSD}                 \\ \hline
System  & \SYS            & RocksDB         & LMDB    & \SYS             & RocksDB         & LMDB    \\ \hline
mean    & \textbf{0.0735} & \textbf{0.1312} & 2.4099 & \textbf{0.2184}  & \textbf{0.2526} & 2.2824  \\
P99     & \textbf{0.7923} & \textbf{0.6364} & 13.799 & \textbf{1.6543}  & \textbf{2.2387} & 12.557  \\
P999    & \textbf{3.0133} & \textbf{3.5250} & 17.794 & \textbf{5.0363}  & \textbf{5.4436} & 16.698  \\ \hline
\end{tabular}
}
\end{table}

\begin{figure*}[htp]
\centering
\includegraphics[width=2\columnwidth]{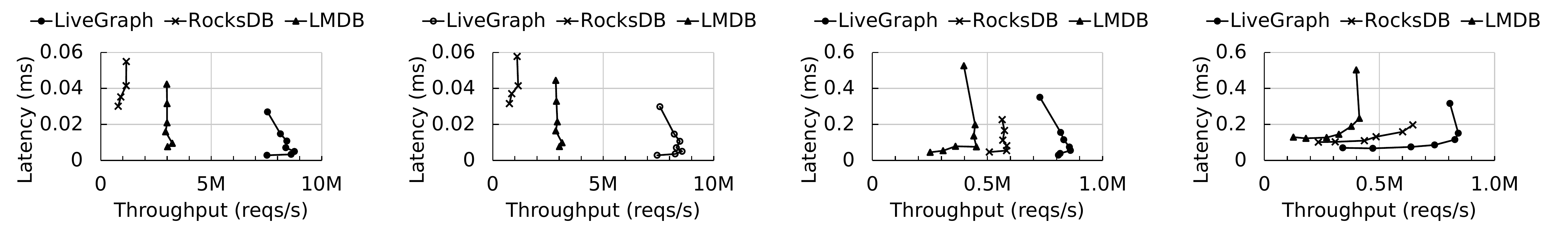}
\begin{subfigure}[b]{0.5\columnwidth}
	\centering
    \caption{Optane in memory \label{fig:tao_inmemory_optane}}
\end{subfigure}
~
\begin{subfigure}[b]{0.45\columnwidth}
	\centering
    \caption{SSD in memory \label{fig:tao_inmemory_ssd}}
\end{subfigure}
\begin{subfigure}[b]{0.45\columnwidth}
	\centering
    \caption{Optane out of core \label{fig:tao_outcore_optane}}
\end{subfigure}
~
\begin{subfigure}[b]{0.45\columnwidth}
	\centering
    \caption{SSD out of core \label{fig:tao_outcore_ssd}}
\end{subfigure}
\caption{TAO throughput and latency trends, with different number of clients and memory hardware}
\label{fig:tao_latency_throughput}
\end{figure*}
\begin{figure*}[htp]
\centering
\includegraphics[width=2\columnwidth]{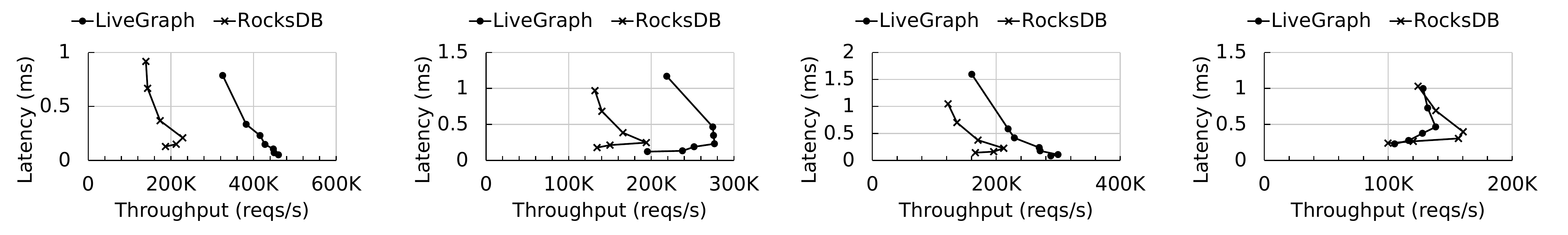}
\begin{subfigure}[b]{0.45\columnwidth}
	\centering
    \caption{Optane in memory \label{fig:dflt_inmemory_optane}}
\end{subfigure}
~
\begin{subfigure}[b]{0.45\columnwidth}
	\centering
    \caption{SSD in memory \label{fig:dflt_inmemory_ssd}}
\end{subfigure}
\begin{subfigure}[b]{0.45\columnwidth}
	\centering
    \caption{Optane out of core \label{fig:dflt_outcore_optane}}
\end{subfigure}
~
\begin{subfigure}[b]{0.45\columnwidth}
	\centering
    \caption{SSD out of core \label{fig:dflt_outcore_ssd}}
\end{subfigure}
\caption{DFLT throughput and latency trends, with different number of clients and memory hardware}
\label{fig:dflt_latency_throughput}
\end{figure*}

\para{Out-of-core latency } Tables~\ref{tab:tao_latencies_ooc} and \ref{tab:linkbench_latencies_ooc} list the out-of-core (OOC) results, enabled by limiting memory (using Linux cgroup tools) to 4GB, which is about 16\% of \SYS's, 9\% of LMDB's and 28\% of RocksDB's memory usage.
This cap is the minimal memory for RocksDB to run with 128 client threads while delivering its peak throughput.

RocksDB is optimized for OOC writes, by dumping sorted blocks of data sequentially and performing compression for better I/O bandwidth usage.
\SYS's design prioritizes reads instead.
It performs sequential writes within an adjacency list but it does not ensure sequential storage of multiple dirty adjacency lists.
It also issues smaller I/O requests by doing page write-back, with write size starting at 4KB, as opposed to the several MBs of RocksDB's LSMT.
Fortunately, low-latency SSDs like our Optane device or byte-addressable NVM alleviate such problems.

Test results confirm the rationale above.
With the more read-heavy TAO, \SYS wins across the board, cutting average latency by 2.19$\times$ from the runner-up LMDB on Optane.
On NAND SSD, RocksDB beats LMDB, being more bandwidth-efficient with its compression.
Still, its average latency is 1.46$\times$ higher than \SYS.
For DFLT, \SYS outperforms RocksDB by 1.15$\times$ on NAND SSD, and by 1.79$\times$ on the faster Optane SSD.


Across the latency tests, by checking the top 2 finishers, it becomes apparent that among existing solutions, the B+-tree-based LMDB and LSMT-based RocksDB offer good performance under mostly-read and mixed workloads, respectively.
However, when placed under unfavorable conditions, they switch places, with a 2$\times$ to 10$\times$ performance gap in between for each latency category.
\SYS, in contrast, provides clearly better performance both in memory and out of core with Optane SSD or with the TAO workload.

\begin{figure}[bt]
\centering
\begin{subfigure}[b]{0.48\columnwidth}
    \includegraphics[width=\columnwidth]{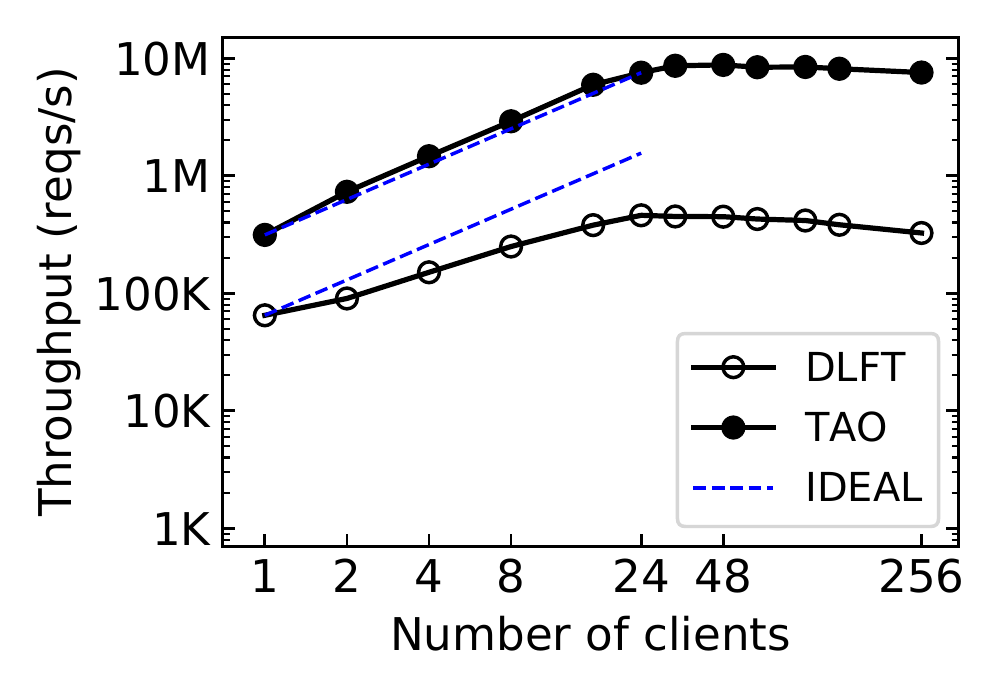}
    \caption{\SYS scalability}
    \label{fig:scalability_oltp}
\end{subfigure}
\begin{subfigure}[b]{0.48\columnwidth}
    \includegraphics[width=\columnwidth]{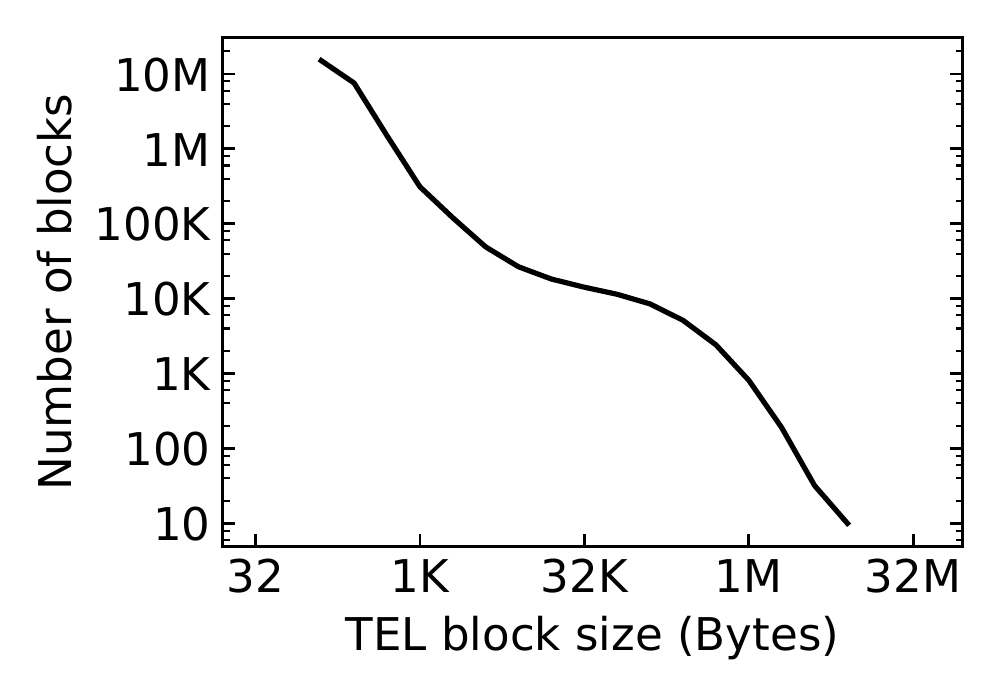}
    \caption{Block size distribution}
    \label{fig:block-dist}
\end{subfigure}
\caption{\SYS scalability and block size distribution}
\label{fig:scalability_block-dist}
\end{figure}

\para{Scalability and throughput.}
We examine the multi-core scalability of \SYS by running LinkBench with an increasing number of clients.
Figure~\ref{fig:scalability_oltp} gives the result.
We can see that \SYS's throughput scales smoothly with more cores being used until 24 clients when all the physical cores are occupied. 
For TAO, the scalability curve is very close to the ideal one.
For DFLT, the write-ahead-logging bottleneck makes it hard for \SYS to achieve perfect scalability.
We expect the emergence of NVM devices and related logging protocols~\cite{arulraj2015let,arulraj2016write} would resolve this issue.

We then saturate the systems to measure throughput under the two workloads, removing the think time between requests. Figures~\ref{fig:tao_latency_throughput} and \ref{fig:dflt_latency_throughput} show latency and throughput when increasing the number of clients from 24 (the number of cores in our server) until the peak throughput is reached, which required up to 256 clients.

For TAO (Figure~\ref{fig:tao_latency_throughput}), when in-memory on top of Optane (for durability), LMDB saturates at 32 clients (i.e., near to the number of physical cores) with a throughput of 3.24M requests/s, after which the contention on the single mutex intensifies.
\SYS's throughput peaks at 8.77M requests/s with 48 clients, strained by context switch overhead afterwards. We get similar results with NAND.

Out of core, running TAO on top of Optane, RocksDB beats LMDB and reaches the peak point at 48 clients with a throughput of 584K requests/s, \SYS reaches 860K requests/s at 64 clients.
With NAND, \SYS still improves the throughput by 1.31$\times$ from RocksDB.

For DFLT (Figure~\ref{fig:dflt_latency_throughput}), RocksDB reaches 228K requests/s in memory and saturates at 48 clients, when compaction starts to pause writes frequently and write/read amplification becomes more significant.
By contrast, \SYS is able to push beyond 460K at 24 clients, as \LOG does not have such limits. NAND SSD results are similar, showing \SYS 4.83$\times$ and 1.43$\times$ faster than the runners-up, respectively.
Out of core with Optane, \SYS peaks at 300K requests/s with 32 client threads and RocksDB saturates at 212K with 48 clients.
With NAND, \SYS reaches 96.1\% of RocksDB's peak throughput.

\para{Memory consumption.}
Using our default compaction frequency every 65536 transactions,  the DFLT workload, 24 client threads, and 12M transactions total,
\SYS consumed 24.9GB in total, 
against 
44.8 GB and 14.4 GB for LMDB and RocksDB,
respectively.
For \SYS, 706 MB space is recycled but not yet used at the end of the DFLT run.
Of the allocated space, the aggregate over-provisioned space is about 4.6GB, leading to 81.2\% final occupancy. 

Figure~\ref{fig:block-dist} gives the \LOG block count distribution at different sizes for this workload, which matches the power-law degree distribution among vertices~\cite{faloutsos1999power}, validating \LOG's ``buddy-system'' design. 

\para{Effectiveness of compaction.}
When compaction is completely turned off, \SYS's footprint sees a 33.7\% increase, requiring 33.3GB space.
Compaction only scans a small dirty set so its time overhead is fairly small: varying the compaction frequency brings insignificant changes in performance ($<$5\%).

\begin{figure}[bt]
\centering
\begin{subfigure}[b]{0.48\columnwidth}
    \includegraphics[width=\columnwidth]{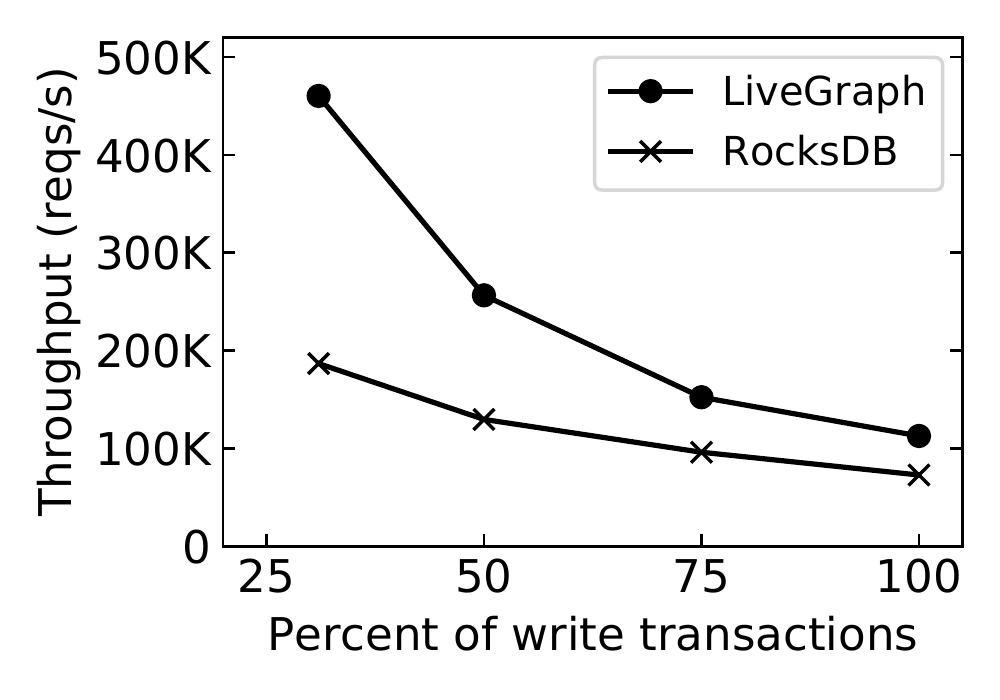}
    \caption{In memory}
    \label{fig:write_intensive_im}
\end{subfigure}
\begin{subfigure}[b]{0.48\columnwidth}
    \includegraphics[width=\columnwidth]{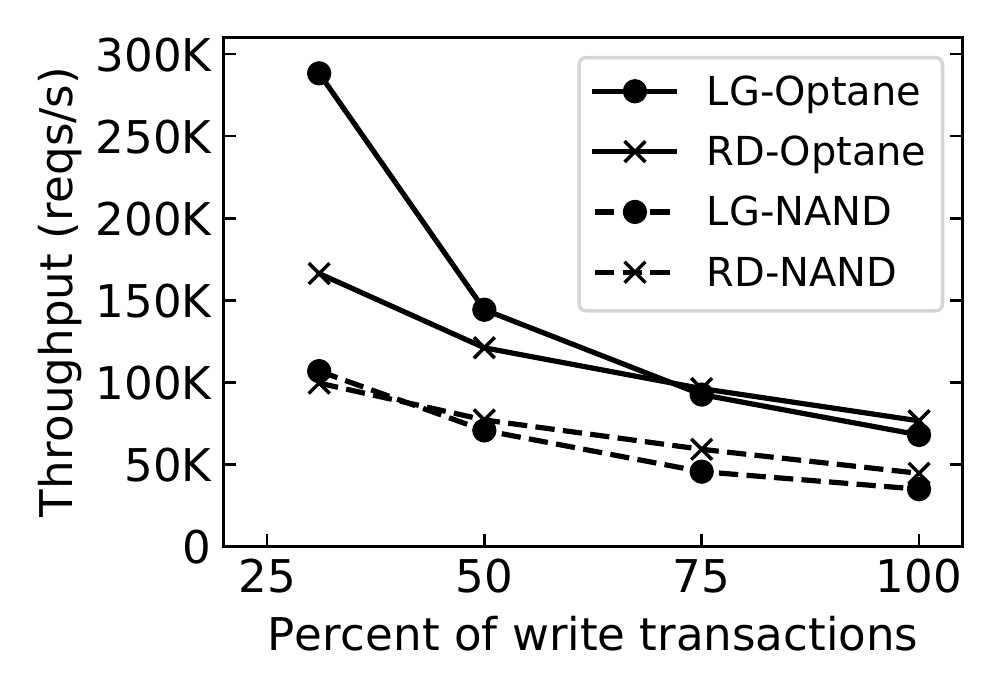}
    \caption{Out of core}
    \label{fig:write_intensive_ooc}
\end{subfigure}
\caption{LinkBench throughput with varying writing ratio}
\label{fig:write_intensive}
\end{figure}

\para{Write-intensive workloads.}
To test more write-intensive workloads, we scale the write ratio, starting from LinkBench DFLT's 31\% to 100\%, and use 24 clients.

Figure~\ref{fig:write_intensive_im} plots the in-memory throughput of LiveGraph and RocksDB (winners of DFLT according to Tables~\ref{tab:tao_latencies_im}-\ref{tab:linkbench_latencies_ooc}), with Optane SSD. 
It shows that \SYS's advantage weakens as the write ratio grows, but still significantly outperforms RocksDB (113K over 73K requests/s, at 1.54$\times$) even with a \textit{write-only} workload.

Figure~\ref{fig:write_intensive_ooc} gives the out-of-core results on both SSDs. 
Here due to the write amplification and random write patterns on disk when out of core, \SYS is overtaken by RocksDB at write ratios of 75\% (Optane SSD) and 50\% (NAND SSD).
Considering that \SYS's design favors read-intensive workloads, which are common in real-world graph database applications~\cite{bronson2013tao,armstrong2013linkbench}, it fares reasonably well with an uncommon, extremely write-intensive workload.
At 100\%-write, \SYS reaches 88.8\% of RocksDB throughput with Optane (78.3\% with NAND).

We also profile disk output volume to measure write amplification and compare with RocksDB, which compresses SSTs to reduce disk I/O. 
\SYS writes 3.02$\times$ as much data as RocksDB: while the latter flushes only sorted updates, the former performs random writes of entire dirty pages and writes larger updates. 
However, the bandwidth of both our SSD devices is larger than the maximum bandwidth used by \SYS (522 MB/s), nearly removing the performance impact of write amplification.
Given the common low write-ratio in commercial workloads, increasing SSD capacities, and falling storage price per GB, we argue that \SYS provides much improved latency and throughput at a small cost for average use cases.

\para{Long-running transactions and checkpoints.}
Our design supports low overhead read-only transactions with snapshot isolation. 
In the following test, we let a checkpointer continuously scan the whole graph and dump the latest snapshot to the NAND SSD, while simultaneously running LinkBench DFLT (in-memory with Optane).

Dumping a snapshot with a single thread to shared-memory takes 16.0s, without concurrent LinkBench queries.
With concurrent queries with 24 clients (one per core), it takes 20.6s (22.5\% slower). 
Meanwhile, LinkBench throughput only slows down by 6.5\% when running concurrent, continuous checkpointing. 
We get similar results on NAND SSD, with 10.9\% and 3.6\% slowdown in checkpointing and LinkBench processing, respectively.
Users may choose to configure \SYS to use more threads for faster checkpointing, at the cost of higher interference.
E.g., 
with 24 checkpointing threads, \SYS completes a 12GB checkpoint in 6.5s, fully utilizing the NAND SSD write bandwidth (2GB/s), while still delivering 86.4\% of the full LinkBench throughput (the one obtained without checkpointing).

\subsection{Real-Time Analytics Workloads}
\label{subsec:snb-perf}

\begin{table}[t]
\center
\caption{Throughput w.\ SNB in memory (reqs/s). $^{(*)}$ For DBMS T we extrapolated an optimistic estimate.}
\label{tab:snb_im}
\resizebox{1.0\columnwidth}{!}{
\begin{tabular}{c|cccc}
\hline
System          & LiveGraph  & Virtuoso & PostgreSQL & DBMS T \\ \hline
Complex-Only    & 9,106   & 292   & 3.79       & 59.3$^{(*)}$     \\
Overall         & 9,420   & 259   & 52.4      & $-$        \\ \hline
\end{tabular}
}
\end{table}

\para{Real-time analytics.} Table~\ref{tab:snb_im} gives the throughput measured from \SYS, Virtuoso,  PostgreSQL, and DBMS T in memory, with only complex reads (referred to as Complex-Only) and all three categories of requests (referred to as Overall), using 48 client and 48 worker threads and Optane SSD for persistence. 
DBMS T has its own SNB driver that only runs requests sequentially.
We extrapolate throughput by running it with one worker and then optimistically assume that the system scales linearly to 48 workers.
The Overall workload uses SNB's official mix: 7.26\% complex queries, 63.82\% short queries, and 28.91\% updates.
RocksDB and LMDB are skipped as (1) these two K-V stores do not have official SNB implementations, and (2) our earlier micro-benchmark and LinkBench results of their basic graph database operations (upon which more complex queries are built) significantly lags behind \SYS.
We only report DBMS T's Complex-Only results as its implementation for SNB update requests is still absent.
DBMS T is single-writer so it is unlikely that it would outperform the other baselines on update requests.
\SYS outperforms the runner-up, Virtuoso, by 31.19$\times$ and 36.43$\times$, producing gains far larger than those observed in microbenchmarks or LinkBench.
Meanwhile, Virtuoso beats DBMS T by 4.92$\times$ and beats PostgreSQL by 77.05$\times$ and 4.94$\times$.


We found fast edge scans are even more critical with complex analytics workloads. 
This also explains why PostgreSQL underperforms, since it does not support clustered indexes~\cite{clustered_index}.
Second, MVCC is crucial for fast analytics when these complex queries are mixed with write transactions. 
Compared to Complex-Only, Overall has more short queries/updates, so \SYS and PostgreSQL both produce higher throughput (the reason for \SYS's small Complex-Only vs.\ Overall difference is that Overall is more write-intensive and throughput is limited by persisting the WAL).
Virtuoso, on the other hand, performs worse by spending over 60\% of its CPU time on locks.  
Of course, MVCC comes with its space overhead: 
for this workload, \SYS consumes about 30GB, PostgreSQL 19GB, and Virtuoso only
8.3GB. 
Compared to PostgreSQL, which also performs MVCC, \SYS's space overhead comes from its longer timestamps, plus its overprovisioning in adjacency lists (a key design choice that helps in 
query performance). 

One may argue that Virtuoso's smaller memory footprint helps with out-of-core execution.
Table~\ref{tab:snb_ooc} (with 3GB DRAM cap, persisting on Optane) shows a heavy performance hit for both \SYS and Virtuoso 
when going out-of-core, and the gap between them does shrink. 
However, \SYS is still an order of magnitude better, and for the Overall mix, beats Virtuoso's \textit{in-memory} performance by $1.35\times$.

Again, though results are not listed, Neo4j and DBMS S lose to all three systems by another 2-3 orders of magnitude. 
This is in part due to their use of Java, which is not ideal for data-intensive workloads, and in part because of their choice of data structures: Neo4j uses linked list and DBMS S builds on copy-on-write B+ trees similar to LMDB.
Our examination reveals that multi-hop graph queries dramatically stress Java's garbage collection. 


\begin{table}[t]
\center

\caption{Throughput w.\ SNB out of core (reqs/s)}
\label{tab:snb_ooc}
\begin{tabular}{c|cc}
\hline
System          & LiveGraph & Virtuoso \\ \hline
Complex-Only    & 31.0     & 2.91     \\
Overall             & 350    & 14.7    \\ \hline
\end{tabular}
\bigskip

\caption{Average latency of queries in SNB (ms)}
\label{tab:snb_lat}
\resizebox{1.0\columnwidth}{!}{
\begin{tabular}{c|cccc}
\hline
System              & LiveGraph & Virtuoso  & PostgreSQL & DBMS T    \\ \hline
Complex read 1      & 7.00      & 23,101 & 371 & 717       \\
Complex read 13     & 0.53      & 2.47      & 10,419 & 19.4     \\
Short read 2        & 0.22      & 3.11      & 3.31 & 3.22          \\
Updates             & 0.37      & 0.93      & 2.19 & $-$         \\ \hline
\end{tabular}
}
\end{table}

\para{Query case study.}
Due to space limits, we present brief case studies on selected queries in the SNB mixes, with their average latencies listed in Table~\ref{tab:snb_lat}.

``Complex read 1'' accesses many vertices (3-hop neighbors). 
Here the benefit of MVCC stands out, with Virtuoso suffering lower throughput caused by locks.

``Complex read 13'' performs pairwise shortest path (PSP) computation. 
Virtuoso's implementation uses its custom SQL extension, with well-optimized PSP primitive, but still loses to \SYS by 4.68$\times$. 
PostgreSQL is limited to only recursive SQL for PSP and loses to other approaches by orders of magnitude. 
This demonstrates the benefit of having a graph specific query language or extensions.

``Short read 2'' is a 1-hop query with many short neighborhood operations.
Here latency relates to seek performance, demonstrated to be \SYS's forte in the microbenchmark results.

``Updates'' shows average latencies of all requests containing updates. 
Though \SYS is designed mainly toward fast reads, it performs well in writes too, leading Virtuoso by 2.51$\times$ and PostgreSQL by 5.92$\times$. 

\subsection{Iterative Long-Running Analytics}
Finally, we evaluate \SYS's capability of performing iterative long-running analytics directly within its primary graph store. 
Table~\ref{tab:snb_olap} presents its performance in running two common graph algorithms, PageRank and Connected Components (ConnComp), on a subgraph (with 3.88M edges) of the SNB SF10 dataset, involving all the Person nodes and their relationships.

\begin{table}
\center
\caption{ETL and execution times (ms) for analytics}
\label{tab:snb_olap}
\begin{tabular}{c|cc}
\hline
System & LiveGraph & Gemini\\
\hline
ETL & - & 1520 \\
\hline
PageRank & 266 & 156 \\
ConnComp & 254 & 62.6 \\
\hline
\end{tabular}
\end{table}

\SYS performs iterative analyses directly on the latest snapshot, 
finishing both tasks under 300 milliseconds, using 24 threads.
For comparison, we give the results of Gemini~\cite{zhu2016gemini}, a state-of-the-art graph engine dedicated to such static graph analytics.
Without the compact yet immutable CSR-based graph storage used by Gemini, \SYS manages to reach 58.6\% and 24.6\% of Gemini's performance for PageRank and ConnComp, respectively. 
In addition, to enjoy Gemini's superior performance, one has to export to its data format, then load the graph into its memory. 
We measured this ETL overhead (converting from TEL to CSR) for this specific graph to be 1520ms, greatly exceeding the PageRank/ConnComp execution time, not to mention the difference between \SYS and Gemini.

\section{Related Work}
\label{sec:relwork}

\spara{Transactional Systems.}
Graph databases can be grouped into two
categories.
Native graph databases~\cite{neo4j,titan,orientdb,arangodb} are designed from scratch for storing/querying graphs.
Non-native graph databases store graphs using general-purpose stores, sometimes using a layer on top of an existing RDBMS or key-value store~\cite{bronson2013tao,sun2015sqlgraph,virtuoso,triad,dubey2016weaver}.

All these systems support transactional graph workloads and some of them support analytical graph workloads~\cite{besta2019demystifying}.
As discussed in the motivation, these systems adopt tree-based~\cite{titan,orientdb,bronson2013tao,sun2015sqlgraph,virtuoso,triad,kimura2017janus,jindal2015graph,dubey2016weaver} or pointer-based~\cite{neo4j} data structures and, therefore, suffer a major limitation of requiring pointer chasing for adjacency list scans.
Even state-of-the-art systems designed for low latency do not optimize for purely sequential scans of adjacency lists~\cite{carter2019nanosecond}.

There are efforts on improving the analytical performance of existing transactional systems~\cite{kimura2017janus,case,jindal2015graph,all-in-one}.
These systems generally compare with and offer comparable performance to GraphLab, while reducing loading time.
However, GraphLab is not a state-of-the-art baseline, as shown by existing literature~\cite{shun2013ligra,nguyen2013galois,zhu2016gemini}.

\spara{Analytics on Dynamic Graphs.}
Several graph engines support graph analytics over an evolving graph, to study graph evolution or to update computation results incrementally.
%
Kineograph~\cite{cheng2012kineograph} supports incremental graph analysis
by periodically applying updates and generating snapshots.
Chronos~\cite{han2014chronos} and ImmortalGraph~\cite{miao2015immortalgraph} are 
designed to analyze graph evolution,
processing a sequence of predefined, read-only graph snapshots.
LLAMA~\cite{macko2015llama} applies incoming updates in batches and creates copy-on-write delta snapshots dynamically
for temporal graph analysis.
Grace~\cite{prabhakaran2012managing} supports transactional updates but uses an expensive copy-on-write technique:
every time an adjacency list is modified, the entire list is copied to the tail of the edge log.
This makes scans purely sequential but it also makes updates very expensive, especially for high-degree vertices.
GraphOne~\cite{kumar2019graphone} serializes edge updates by appending them onto a single edge log.
It does not support transactions or durability.

These systems focus primarily on graph analysis.
\SYS supports real-time transactional workloads with better performance than existing graph databases, while supporting whole-graph analytics on the same primary graph store.
Many incremental analytics techniques above can readily be incorporated, leveraging \SYS's multi-versioning.



\spara{Graph engines.}
Graph engines perform analytical processing, such as graph processing~\cite{gonzalez2012powergraph,gonzalez2014graphx,zhu2016gemini,shun2013ligra,nguyen2013galois,sundaram2015graphmat,xie2013fast} or graph mining~\cite{wang2015graphq,teixeira2015arabesque,chen2018g}.
Their design assumes an immutable graph topology,
hence widely adopting read-optimized CSR/CSC
representations.
As discussed earlier, this delivers superior analytics performance, but does not 
handle
updates/insertions.
Hence existing graph engines have been limited to processing static, stale snapshots dumped from the data source.
In contrast, \SYS supports analytics on dynamic graphs  without costly ETL.

\section{Conclusion and Future Work}
\label{sec:conclusions}
%
%
Our work shows that it is possible to design graph data management systems that are fast at both transactional and analytical workloads.
The key is using data structures that are tailored to the operations of graph workloads, as well as associated algorithms for transactional support.
Our evaluation confirms the strength of \SYS as a potential all-around choice  across multiple graph workloads.

The next steps of \SYS include adapting relational concurrency control and storage techniques optimized for modern hardware~\cite{neumann2015fast,arulraj2016write,lim2017cicada,leis2018leanstore}.
\SYS's design is amenable to scaling out, 
leveraging techniques in distributed graph query processing~\cite{shi2016fast,zhang2017sub,dubey2016weaver} and distributed transaction management~\cite{peng2010large,corbett2013spanner,binnig2014distributed,dubey2016weaver}.
In addition, \SYS can be extended in another direction: the multi-versioning nature of \LOGs makes it natural to support temporal graph processing~\cite{han2014chronos,miao2015immortalgraph}, with modifications to the compaction algorithm to efficiently store and index older graph versions.



\section*{Acknowledgments}
We sincerely thank all our VLDB 2020 reviewers for their insightful comments and suggestions. 
We appreciate suggestions from colleagues Saravanan Thirumuruganathan, Nan Tang, and Mourad Ouzzani during our revision.
This work is supported in part by NSFC Grant 61525202 for Distinguished Young Scholars.

\balance

\bibliographystyle{abbrv}
\bibliography{refs}  





\end{document}